\documentclass[aoas,nameyear,seceqn,dvips]{arximspdf}
\usepackage{graphics}
\doi{10.1214/10-AOAS340}
\volume{4}
\issue{3}
\pubyear{2010}
\firstpage{1365}
\lastpage{1382}


\begin{document}
\begin{frontmatter}

\title{Analysis of spatial distribution of marker expression
in cells using boundary distance plots}
\runtitle{Boundary distance plotting}

\begin{aug}
\author[A]{\fnms{Kingshuk} \snm{Roy Choudhury}\thanksref{T1}\corref{}\ead[label=e1]{kingshuk@ucc.ie}},
\thankstext{T1}{Supported by Research Frontiers Grant 07/MATF/543 from the Science Foundation Ireland.}
\author[B]{\fnms{Limian} \snm{Zheng}}
\and\\
\author[B]{\fnms{John J.} \snm{Mackrill}\thanksref{T2}}
\thankstext{T2}{Supported by Research Frontiers Grant 07/BMIF/548 from the Science Foundation Ireland.}

\runauthor{K. Roy Choudhury, L. Zheng and J. J. Mackrill}
\affiliation{University College Cork}
\address[A]{K. Roy Choudhury\\
Department of Statistics\\
University College Cork\\
Ireland\\
\printead{e1}}
\address[B]{L. Zheng\\
J. J. Mackrill\\
Department of Physiology\\
University College Cork\\
Ireland}
\end{aug}

\received{\smonth{6} \syear{2009}}
\revised{\smonth{1} \syear{2010}}

\begin{abstract}
Boundary distance (BD) plotting is a technique for making
orientation invariant comparisons of the spatial distribution of
biochemical markers within and across cells/nuclei. Marker expression is
aggregated over points with the same distance from the boundary. We
present a suite of tools for improved data analysis and statistical
inference using BD plotting. BD is computed using the Euclidean distance
transform after presmoothing and oversampling of nuclear boundaries.
Marker distribution profiles are averaged using smoothing with linearly
decreasing bandwidth. Average expression curves are scaled and
registered by \textit{x}-axis dilation to compensate for uneven lighting and
errors in nuclear boundary marking. Penalized discriminant analysis is
used to characterize the quality of separation between average marker
distributions. An adaptive piecewise linear model is used to compare
expression gradients in intra, peri and extra nuclear zones. The
techniques are illustrated by the following: (a) a two sample problem
involving a pair of voltage gated calcium channels (Cav1.2 and AB70)
marked in different cells; (b) a paired sample problem of calcium
channels (Y1F4 and RyR1) marked in the same cell.
\end{abstract}


\end{frontmatter}
\section{Introduction}\label{s1}

Optical fluorescence microscopy (OFM) offers a high resolution view of
the morphology and spatial organization of intact cells and organelles.
Various proteins, nucleic acids and metabolites can be individually
labeled with different fluorescent colors, giving an in vivo
picture of their behavior and role in living cells [\citet{fernandez}].
The increasing quality and quantity of these images
necessitate quantitative, indeed statistical, analysis of the inherent
spatial and morphological information. In this paper we consider the
problem of mapping the spatial distribution of marker expression in
reference to distance from the cell/nuclear boundary.

\begin{figure}
\begin{tabular}{@{}cc@{}}
(a)&(b)\\

\includegraphics{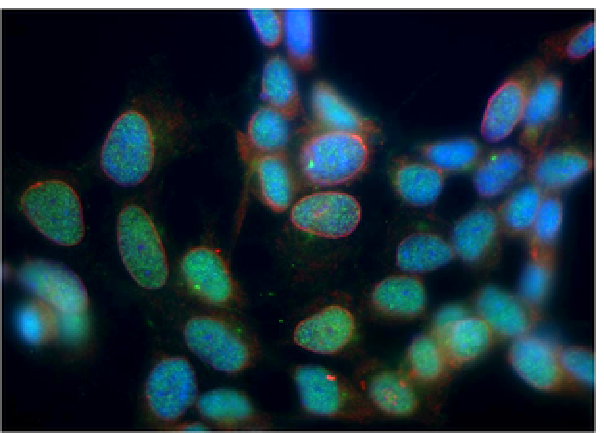}
&\includegraphics{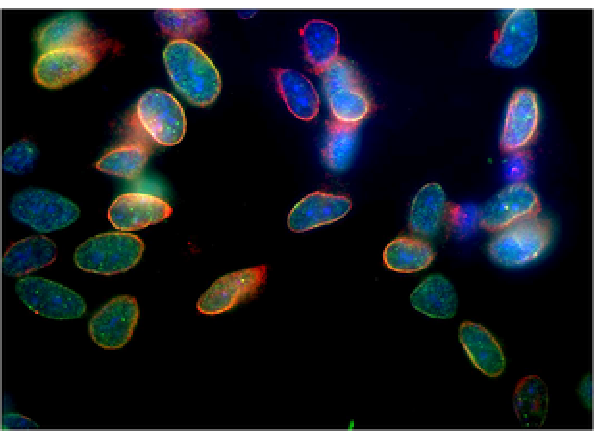}\\
(c)&(d)\\

\includegraphics{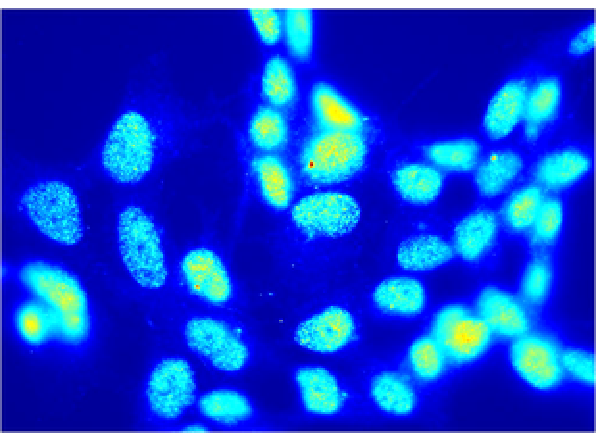}
&\includegraphics{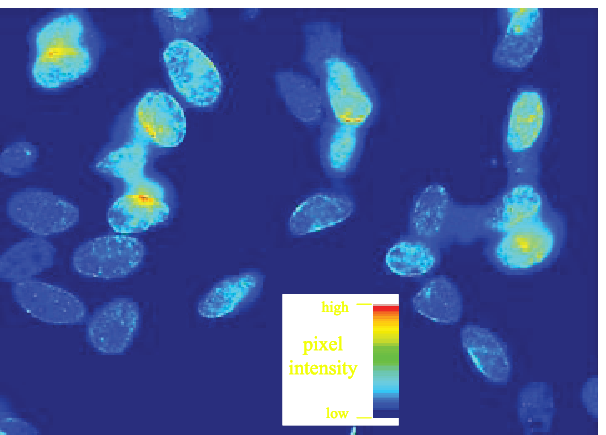}
\end{tabular}
  \caption{Optical fluorescence microscopy image of SH-SY5Y neuroblastoma
cells. Images are $3840\times 3072$ pixels, 8-bit discretization, with
0.08 $\mu\mbox{m}\times 0.08$ $\mu\mbox{m}$ pixel size, acquired using a Nikon Eclipse E600
epifluorescent microscope with $60\times$ objective. Images are labeled in a
blue chromatin marker (`DAPI'), a red nuclear membrane marker (`Emerin')
and \textup{(a)} a green marker (`AB70'), selective for either Cav1.2 or
Cav1.3 VGCC's, \textup{(b)} a green marker (`Cav1.2') selective
for only Cav1.2 VGCC. \textup{(c)}--\textup{(d)} Psuedo-color image of green (AB70)
channel of image in \textup{(a)}. \textup{(d)} Pseudo-color image of green
(Cav1.2) channel of image in \textup{(b)}.}\label{f1}
\end{figure}

First, we consider an experiment comparing the distribution of two
different voltage-gate calcium channels (VGCC), which play an important
role in linking (a) muscle excitation with contraction and (b) neuronal
excitation with transmitter release. New research indicates they may
play a role in gene transcription [\citet{gomez}]. The
VGCC's Cav1.2 and Cav1.3 are studied in the nuclei of the human
neuroblastoma cell-line SH-SY5Y. The images (Figure \ref{f1}) are labeled with
three different fluorescent dyes: (i) a blue chromatin marker, 4,6-diamidino-2-phenylindole (`DAPI'), which essentially marks the body of
the nuclei; (ii) a red mouse monoclonal antibody recognizing the nuclear
matrix protein emerin which lines the nuclear membrane (`emerin'); (iii)
a green antiserum which recognizes either (a) both Cav1.2 and 1.3
(`AB70') or (b) only Cav1.2 (`Cav1.2'). Details of the experiment can be
found in \citet {callinan}. The green markers are used as proxy for
presence of the VGCC. Of particular interest is the proximity of the
VGCC to the nuclear membrane, which can give clues to its role in signal
transmission to/from the nucleus. There appears to be considerable
variability in the green marker distribution both within and across
cells [Figures \ref{f1}(c) and (d)]. This suggests that comparisons across the
images can only be accomplished in a distributional or average sense.
Since the orientation of nuclei is modified arbitrarily during cell
fixation in the slide, any analysis conducted on this data should
ideally be orientation invariant.

As a second example, we compare the distributions of an intracellular
calcium release channel, the type 1 ryanodine receptor (RyR1) with that
of the sarcoplasmic (SR)/endoplasmic reticulum (ER) calcium ATPase
(SERCA), an enzyme that pumps Ca$^{2+}$ into the lumen of intracellular
Ca$^{2+}$ stores such as the SR and ER. In this experiment JEG-3
trophoblastic cells [Figure \ref{f7}(a)] were labeled with the following: (i)
DAPI (blue marker) to mark the body of the nucleus; (ii) a mouse
monoclonal antibody Y1F4 (red marker) recognizing all SERCA subtypes;
and (iii) a rabbit polyclonal antiserum recognizing the type 1 RyR
subtype only (RyR1, green marker). Analysis of the distribution of RyRs
in the trophoblastic cell-line JEG-3 is of interest because the roles of
these calcium channels in nonmuscle cell types, such as these placental
epithelial cells, have not been extensively characterized. In muscle
cells, RyR channels play a pivotal role in coupling extracellular
signals to the release of calcium from the SR/ER, which triggers
activation of the contractile apparatus [\citet{mackrill}]. We anticipate
that RyR1, a channel that releases Ca$^{2+}$ from the SR/ER, would
display a similar distribution to that of SERCA, the main pumping system
that actively accumulates Ca$^{2+}$ into this organelle.

\begin{figure}
\begin{tabular}{@{}cc@{}}
(a)&(b)\\

\includegraphics{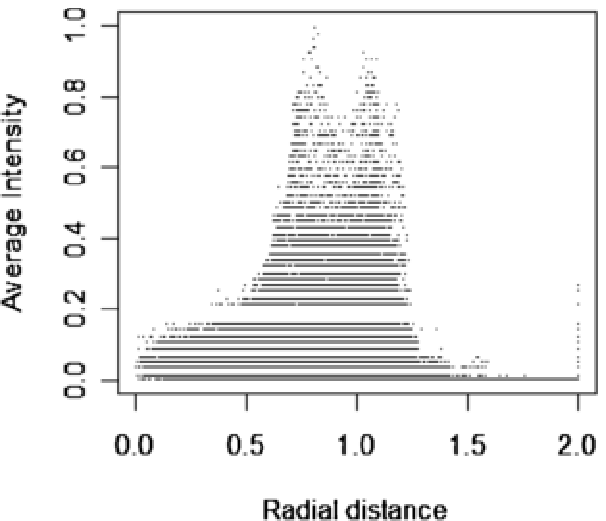}
&\includegraphics{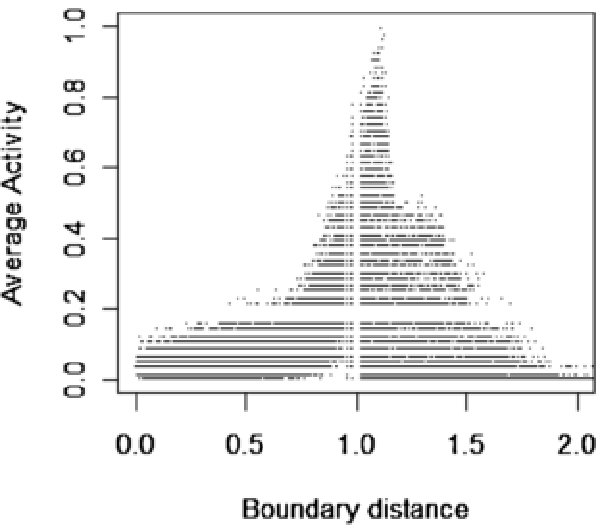}\\
(c)&(d)\\

\includegraphics{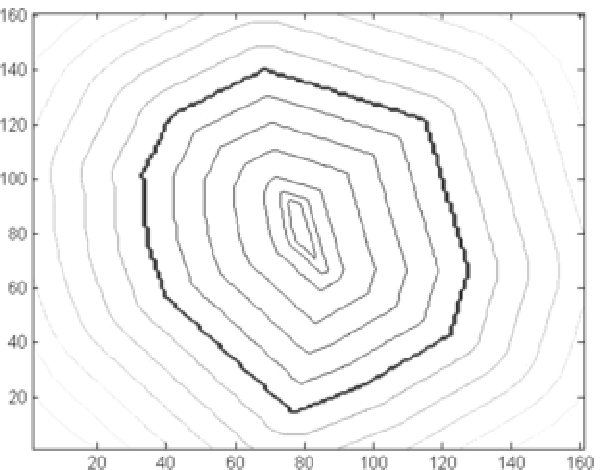}
&\includegraphics{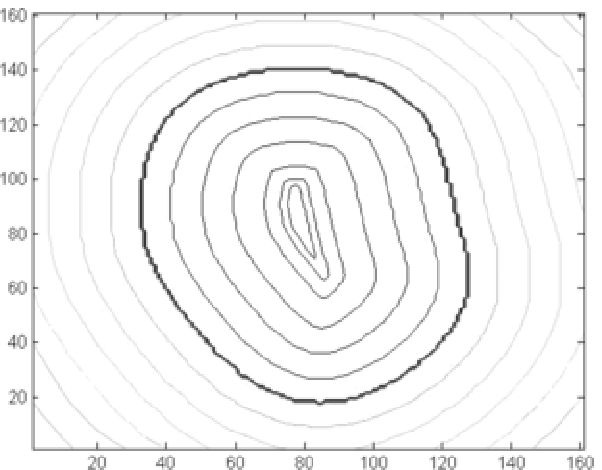}\\
\multicolumn{2}{@{}c@{}}{(e)}\\
\multicolumn{2}{@{}c@{}}{
\includegraphics{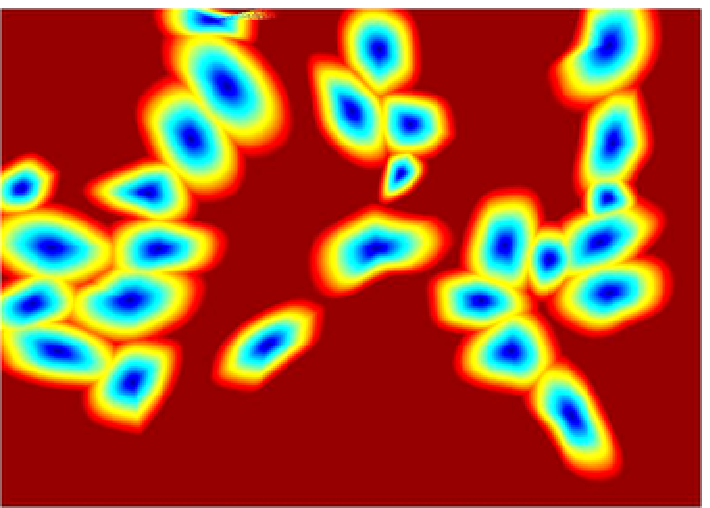}
}
\end{tabular}
  \caption{\textup{(a)} and \textup{(b)} Profile distribution of emerin (red marker) for a
nucleus in Figure \protect\ref{f1}\textup{(b)}. Each point in the plots represents the observed
red channel intensity at a pixel in the image. Expression of point
$(x,y)$ is plotted against \textup{(a)} the radial distance from the
center of the nucleus, \textup{(b)} distance to nearest point on nuclear
boundary. \textup{(c)}--\textup{(e)} Boundary distance (BD) maps \textup{(c)} of the polygonal
region representing a nucleus, \textup{(d)} same nucleus, but smoothed
boundary,
\textup{(e)} pseudocolor image showing orbits for all nuclei in Figure \protect\ref{f1}\textup{(b)}.}\label{f2}
\end{figure}

From a statistical perspective, this problem involves the comparison of
marker expression distributions across cells and experimental
conditions. When orientation is not of interest, it is convenient to
reduce the two-dimensional distribution of markers to one-dimensional
profiles, plotted\break against a common `distance.' This makes it easier to
superimpose and visualize multiple profiles across cells on the same
plot. As each nucleus/cell has a different shape and size, measurement
of proximity must be adapted to the shape or `geometry' of each
individual nucleus/cell. For instance, when we consider distribution of
the boundary marker emerin (red) for a typical nucleus in Figure \ref{f1}(b),
the profile distribution of expression generated by plotting against
radial distance (from the center of the cell) appears to have a bimodal
distribution [Figure \ref{f2}(a)]. By contrast, when we use \textit{boundary}
distance (BD), that is, the distance of each point to the nearest
nuclear boundary (Section \ref{s2}), the profile distribution for the same
nucleus appears to have a single sharp peak at 1 [Figure \ref{f2}(b)]. The
difference is because, unlike radial distance, the level sets of BD are
individually adapted to the nuclear/cell boundary [Figures \ref{f2}(c) and (d)].
BDs can be computed using algorithms such as Euclidean distance mapping
(EDM) [\citet{fabbri}] or morphological erosion [\citet{jahne}].
BDs are normalized to a common scale, for example, 1 at center and 0 at
the boundary, to allow comparison across cells/nuclei of different
shapes and sizes [\citet{knowles}]. In this paper we consider an
extension of EDM to compute BD both within and outside the cell/nucleus.
We also propose the use of oversampled smoothed boundaries for
computation of BD to correct for polygonization of the cell/nuclear
boundary during manual identification (Section \ref{s2}).

Previous analyses of profile distributions from boundary distance plots
have been basically descriptive [\citet{bewersdorf};
\citet{knowles}]. In this paper we develop methods for improved
estimation and statistical inference from profile distributions. For
this purpose, we construct smooth average expression curves to summarize
the profile distribution for each nucleus. In particular, we show why a
linearly increasing bandwidth for smoothing is necessary (Section \ref{s31}).
Variations in light intensity across the image are compensated by
scaling expression curves (Section \ref{s311}). Uneven blue staining near the
boundary of the nucleus can cause incorrect boundary identification,
which was originally thought to affect only distances near the boundary
[\citet{bewersdorf}]. By analyzing this as an errors in variables
type problem, we show that estimated BDs are biased upward. To correct
for this, we realign average expression curves using an \textit{x}-axis dilation
prior to statistical analysis (Section \ref{s312}). Next, we show how methods
such as \textit{t}-tests and penalized discriminant analysis can be used to
describe the differences between groups of profile average expression
curves (Section \ref{s322}). We also use a knot-adaptive piecewise linear
model to draw inferences about expression curves and their derivatives
within regions of interest in the nuclei (Section \ref{s33}). Section
\ref{s4} applies this methodology to the second example. Section \ref{s5} concludes with
a summary of findings and their scientific implication.

The main steps involved in BD analysis are listed below. The sections of
the paper where these steps are described in detail are given in
brackets:

\begin{enumerate}
\item Mark cell/nucleus boundaries and compute BD maps (\ref{s21}).

\item Obtain average marker expression curves for each cell (\ref{s31}).

\item Align activity curves group by scaling and shifting (unpaired: \ref{s311} and \ref{s312}, paired: \ref{s4}).

\item Comparison across groups using functional data analysis (unpaired: \ref{s32}, paired: \ref{s4}).

\item Comparison of expression gradients using piecewise linear models (\ref{s33}).
\end{enumerate}

\section{Computing boundary distances}\label{s2}

If we represent a cell/nucleus as a point set $R$, the Euclidean
distance transform (EDT) of a point $p$ within $R$ is defined as
$D(R^{c},p)=\operatorname{inf}\{d(p,q)\mid q \in R^{c}\}$, that is, the distance of
the $p$ from the nearest point in the complement of $R$. Let $d_{m}=\sup\{D( R^{c},p),p\in R\}$ denote the `maximal distance' from the
boundary. To obtain a scaled BD map that is 0 at the `center' of $R$ and
1 at the boundary and extends continuously for outside $R$, we define
\begin{equation}\label{e21}
\mathrm{BD}(p)=\cases{1 - d_{m}^{ - 1}D( R^{c},p),&\quad$p\in R$,
\cr1 + d_{m}^{ - 1}D( R,p ),&\quad$p\in R^{c}$.}
\end{equation}

A number of efficient algorithms for computing the EDT have appeared
over the last decade or so [\citet{fabbri}] and many of these are
available in standard image analysis packages such as the freely
available ImageJ (\url{http://rsbweb.nih.gov/ij/}). Contours of
the BD function resemble the cell boundary for points near the boundary,
but not necessarily for points deep in the interior [Figures \ref{f2}(c) and
(d)].

\subsection{Boundary smoothing}\label{s21}

To construct the BD, we first need to identify cell/nucleus boundaries
using either automated segmentation methods [\citet{jahne}] or hand
drawing. When segmented regions are polygonal, so too are the contours
of the resulting BD map [Figure \ref{f2}(c)], whereas we know that nuclear
membranes have a much smoother shape. We use periodic smoothing splines
to smooth the boundary curve [\citet{wahba}]. The fitted curve appears to
circumscribe the polygon defined by the original boundary points
[Figures \ref{f2}(c) and (d)]. The fitted curve is sampled at a large number
of points (1000) to generate the smooth boundary. The resulting contours
are typically visually more satisfactory. When BD is computed for an
image with multiple nuclei, a decision rule is required to assign points
to `orbits' of particular nuclei. In Figure \ref{f2}(e) an \textit{orbit} of a
particular nucleus consists of all points whose BD is smallest relative
to distances to other nuclei. Nuclei/cells that lie on the boundary of
the image are ignored from subsequent analysis.

\begin{figure}
\begin{tabular}{@{}cc@{}}
(a)&(b)\\

\includegraphics{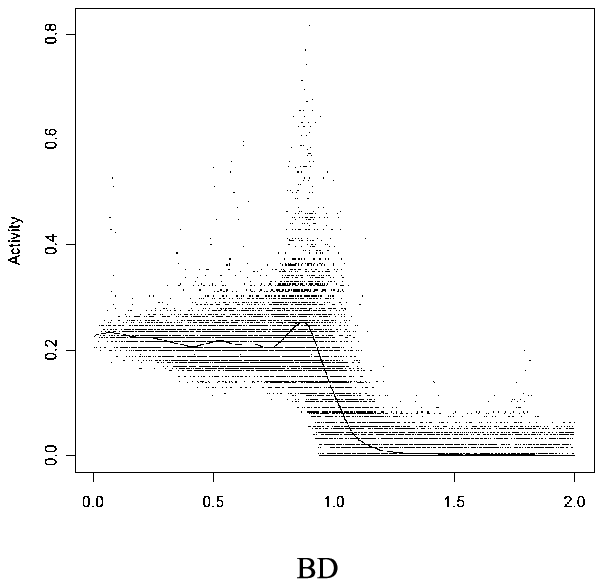}
&\includegraphics{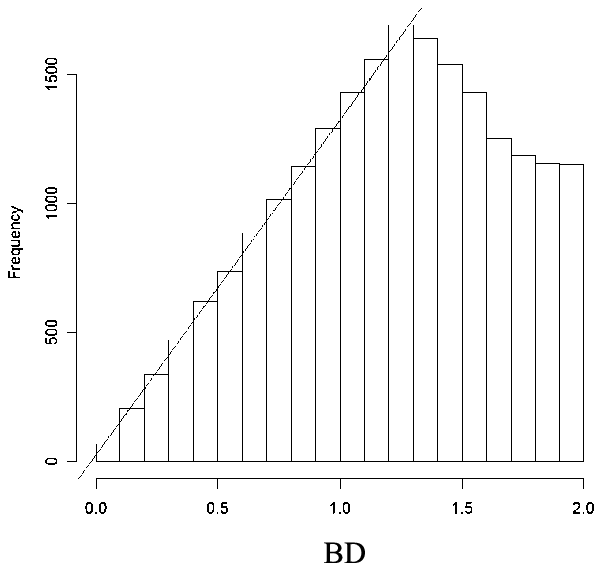}\\
\multicolumn{2}{@{}c@{}}{(c)}\\
\multicolumn{2}{@{}c@{}}{
\includegraphics{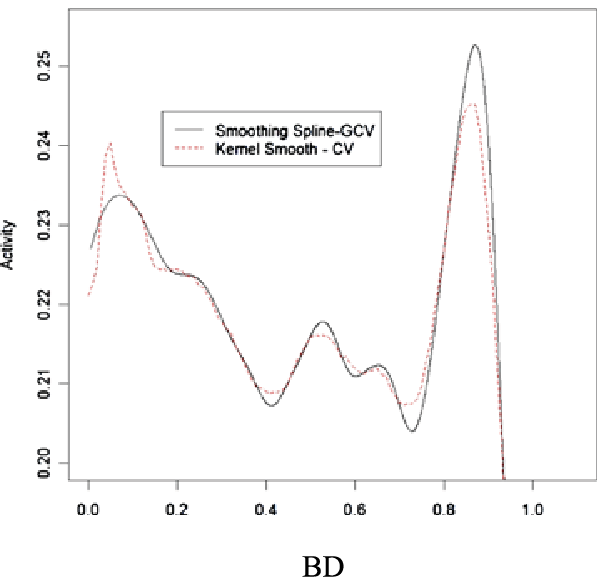}
}
\end{tabular}
  \caption{\textup{(a)} Distribution of green (Cav1.2) expression against
boundary distance (BD) for one nucleus in Figure \protect\ref{f1}\textup{(b)}. Line shows local
average computed by smoothing spline. \textup{(b)} Histogram of number of
pixels sampled as a function of BD for radial plot in \textup{(a)}.
Fitted line shows approximation to linear trend from center to nucleus
boundary. \textup{(c)} Close-up view of local average of expression shown in
Figure \protect\ref{f1}\textup{(b)}, showing differences between variable bandwidth smoothing
spline and fixed bandwidth kernel estimates.}\label{f3}
\end{figure}

\section{Statistical analysis of profile distributions}\label{s3}

Let $h(a|r)$ be the profile distribution of expression $a$ at BD $r$.
For analysis across nuclei, we summarize $h$ by its conditional
expectation $g(r)=\mathrm{E}[h(a|r)]$. We call $g$ the average expression
curve. If we assume $g$ to be a smooth function of $r$, for a wide class
of distributions $h, g$ can be estimated by nonparametric regression of
the point cloud $h(a|r)$ as a function of \textit{r} [Figure
\ref{f3}(a)] [\citet{silverman1985}]. Its computation is described in Section~\ref{s31}. We then
present methods for alignment and analysis of these curves across
nuclei.

\begin{figure}
\begin{tabular}{@{}cc@{}}
(a)&(b)\\

\includegraphics{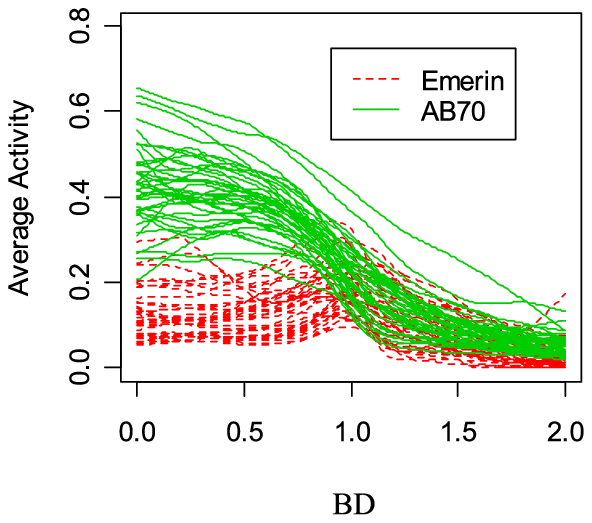}
&\includegraphics{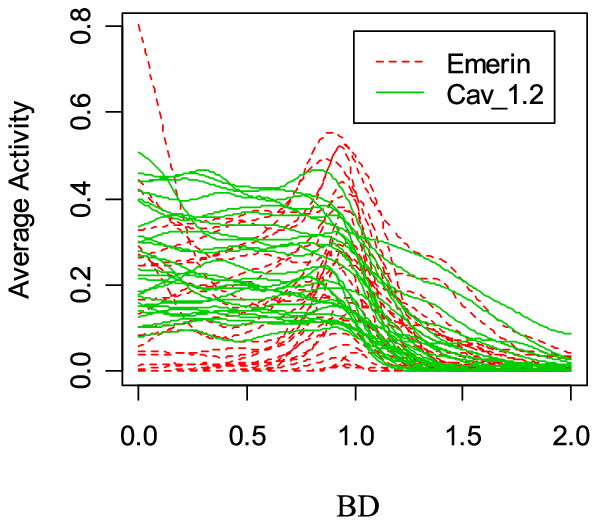}\\
(c)&(d)\\

\includegraphics{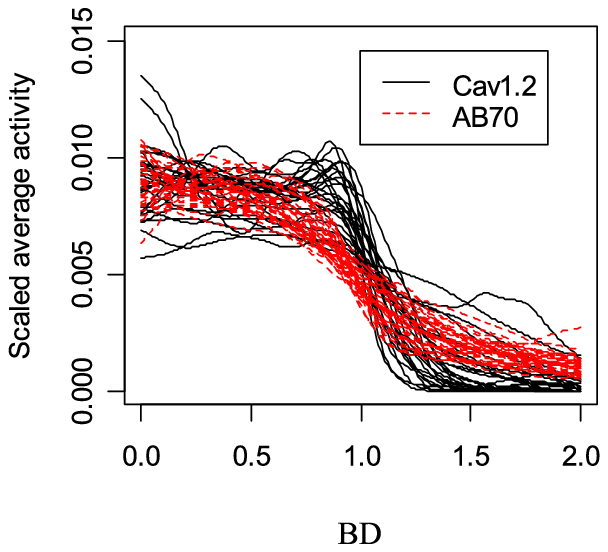}
&\includegraphics{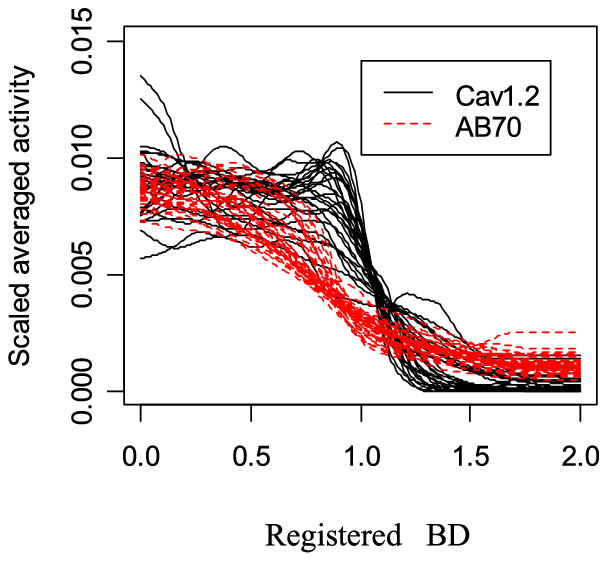}
\end{tabular}
  \caption{Smoothed average expression curves corresponding to:
  \textup{(a)} Nuclei in Figure \protect\ref{f1}\textup{(a)}. Red denotes emerin expression and green denotes
Cav1.2 expression. \textup{(b)} Nuclei in Figure \protect\ref{f1}\textup{(b)}. Red denotes
emerin expression and green denotes AB70 expression. \textup{(c)} Scaled
expression curves corresponding to Cav1.2 (solid black lines) and
AB70 (dashed red lines) expression. The area under each curve \mbox{equals 1}.
\textup{(d)} Scaled expression curves corresponding to Cav1.2 (solid
black lines) and AB70 (dashed red lines) plotted against registered BDs.
Curve registration was done by individual dilation of boundary BDs.}\label{f4}
  \end{figure}

\subsection{Estimating average expression curves}\label{s31}

Assuming that the nucleus is approximately elliptical, its contours have
circumference $0.5\pi er$, where $r$ is length of the minor axis and $e$
is the eccentricity of the ellipse. Thus, the number of points on a
given constant BD contour increases linearly as a function of BD.
Although the nuclei are not exactly elliptical, the accuracy of this
approximation for a typical nucleus in Figure \ref{f1}(b) is empirically borne
out in Figure \ref{f3}(b). For distances beyond the boundary $(r=1)$, this
relationship need not hold, as orbits may compete for points.

In situations where the density of points, $f(r)$, is not
constant,\break
smoothing with a fixed bandwidth can pose problems. In\break Figure \ref{f3}(c), the
fixed bandwidth Nadaraya--Watson estimator $\hat{g}^{\mathrm{N}\mbox{--}\mathrm{W}}( r ) = \{
\sum_{i} K( b^{ - 1}( r - r_{i} ) ) \}^{ - 1}\{ \sum_{i} K( b^{ - 1}( r
- r_{i} ) )h( a_{i}|r_{i} ) \}$ appears to be more variable near the
center (low density) and have more bias near toward the boundary (high
density) [Figure \ref{f3}(c)]. By contrast, the `optimal' variable bandwidth
kernel smoother has bandwidth proportional to $f(r)^{-0.2}$ [\citet{silverman1984}]. The smoothing spline estimator $\hat{g}(r)=\operatorname{arg\min}\{\sum_{i}(h(a_{i}|r_{i})-g(r_{i}))^{2}+\lambda\int\{
g''( u ) \}^{2}\,du\}$ has an equivalent `bandwidth'
proportional to\break $f(r)^{-0.25}$, that is, almost optimal [\citet{silverman1984}]. It appears to produce a more satisfactory estimate [Figure
\ref{f3}(c)]. We will therefore use it for estimating average expression
curves, with $\lambda$ being chosen by generalized cross-validation. For
ease of further computation, each estimated average expression curve is
evaluated at a common grid of regularly spaced points $r_{i} = 0.01i$, $i=1,2,\ldots,200$, in the interval $(0,2]$.

\subsubsection{Scaling expression curves}\label{s311}

In Figures \ref{f4}(a) and (b) we see that the average expression curves of
emerin (red marker) peak near the boundary, whereas the average
activities of Cav1.2 and AB70 are high inside the nuclei and low beyond
the boundary. However, there appears to be a lot of variation in the
amplitudes/scales of these curves. We define scale $s_{k}=\int_{0}^{2}g_{k}(r)\,dr$, where $g_{k}$ is the average expression curve for the
$k$th cell: it is approximated as a Riemann sum, $\hat{s}_{k}=\sum\hat{g}_{k}( r_{i} )$. There appears to be moderate to strong
positive correlation between the red and green scale values across cells
within each image: for Cav1.2 and Emerin, $\hat{\rho} = 0.77$ and for
AB70 and Emerin, $\hat{\rho} = 0.69$. This suggests that at least some of
the variation in scaling may be due to uneven illumination across the
image, affecting both red and green channels. If, on the other hand,
there had been fluoro marker sensitive effects like photo bleaching, it
would most likely affect a single channel locally, producing poor
correlation. To eliminate this extraneous source of variation, we
normalize the curves $g_{k}$ by dividing by the factor estimated scale
factor $\hat{s}_{k}$. The scaled profiles now all subtend an area of 1:
they give us the average `distribution' of the marker in each nucleus as
a function of distance. The distribution curves [Figure \ref{f4}(c)] show much
less intra group variation in the \mbox{$y$-direction} than the unscaled
versions [Figures \ref{f4}(a) and (b)].

\subsubsection{Dilation based registration}\label{s312}

Uneven DAPI staining leads to errors in identification of the true
nuclear boundary. Assuming additive measurement errors in true boundary
distances $D( R^{c},p )$, we can write expected value of the resulting
observed boundary distance BD$_{o}(p)$ as follows:
\begin{eqnarray}\label{e31}
\mathrm{E}[\mathrm{BD}_{o}(p)]
&=&
\mathrm{1-E}\bigl[(d_{m}+\varepsilon _{m})^{-1}\bigl(D(R^{c},p)+\varepsilon_{p}\bigr)\bigr]\nonumber
\\[-8pt]\\[-8pt]
&=&
\mathrm{1-E}\bigl[(1+d_{m}^{-1}\varepsilon_{m}
)^{-1}\bigl(1-\mathrm{BD}(p)+d_{m}^{-1}\varepsilon_{p}\bigr)\bigr]'.\nonumber
\end{eqnarray}

We further assume that the measurement errors $\varepsilon_{p}$ and
$\varepsilon_{m}$ are i.i.d. U[$-\mathit{e},\mathit{e}$]. Taking expectations with
respect to the uniform distribution, we get
\begin{eqnarray}\label{e32}
\mathrm{E}[\mathrm{BD}_{o}(p)]
&=&
1-\bigl(\ln(1+d_{m}^{-1}e)-\ln(1-d_{m}^{-1}e)\bigr)\bigl(1-\mathrm{BD}(p)\bigr)\nonumber
\\[-8pt]\\[-8pt]
&\approx&
1-(1+3d_{m}^{-
2}e^{2})\bigl(1-\mathrm{BD}(p)\bigr)\approx(1+3d_{m}^{-2}e^{2})\mathrm{BD}(p).\nonumber
\end{eqnarray}

The first approximation in (\ref{e32}) follows from a first order Taylor series
expansion. The second approximation assumes $e\ll\mathrm{BD}(p)$. Thus
(\ref{e32}) shows that estimated BD have an upward bias, which can be modeled by a
location independent scale factor.

In terms of observed boundary distances, we can thus write a model for
the observed expression as $z_{k}(p)=g_{k}(\delta_{k}\mathrm{BD}_{o}(p))+\varepsilon(x,y)$. Differential dilation of nuclear boundaries causes
misalignment of expression curves across nuclei. To realign them, we
first estimate the parameters $\delta_{k}$ by minimizing the within
image registration sum of squares:
\begin{equation}\label{e33}
\mathit{WREGSSE}=\sum_{k = 1}^{nc}\int_{0}^{2}w(r)\bigl(g_{k}(r\delta_{k})-\mu(r)\bigr)^{2}\,dr.
\end{equation}
Here $\mu$ is the (unknown) mean curve across nuclei within the group
(Cav1.2 or AB70), $nc$ is the number of nuclei in each image ($=$27
for Cav1.2, $=$38 for AB70) and $w$ is a weighting function which
reflects the precision of the estimated curves (see Section \ref{s31}). Here
$w(r)=r^{0.75}$, $0 < r < 1$; $w(r) = 1$, $0 < r < 1$; $w(r) = 0$ otherwise,
based on the fact that the variability of the smoothing spline estimate
is proportional to $f(r)^{-0.75}$ [\citet{silverman1985}], where $f(r)$ is
the density derived in Section \ref{s31}. Minimization of \textit{WREGSSE} can
be achieved through a two-step Procrustes type iterative procedure
[\citet{ramsay}]. Step 1: The group mean $\mu$ is estimated
by the sample mean of the scaled expression curves $g_{k}$. Step 2:
Given $\mu$, the criterion (\ref{e33}) is separable in the $\delta_{k}$, each of
which can be estimated by a line search procedure. We start with an
initial estimate of \mbox{$\delta_{k}=0$} $\forall k$ and steps 1 and 2 are
iterated to convergence. In this case, iterating only 1~step of the
iterative algorithm resulted in a 15\% reduction of \textit{WREGSSE} for
the Cav1.2 image and 8\% for the AB70 image. Further steps did not
result in any significant decrease of \textit{WREGSSE}. The registered
curves $g_{k}(r+\hat{\delta}_{k})$ appear to exhibit greater
location alignment [Figure \ref{f4}(d)]. Within image registration yields
estimated group mean curves, $\hat{\mu}_{C}$ and $\hat{\mu}_{A}$ for
Cav1.2 or AB70 respectively. These are then registered to each other by
minimizing the sum of squares difference, $\mathit{BREGSSE}=\int_{0}^{2}(\hat{\mu}_{A}(r\delta_{A})-\hat{\mu}_{C}(r))^{2}\,dr$. The
parameter $\delta_{A}$ is estimated by line search and results in a 53\%
reduction in \textit{BREGSSE}. The combined dilation for nuclei in the
AB70 group is thus given by the product $\delta_{A}\delta_{k}$.

\subsection{Functional data analysis (FDA)}\label{s32}

\subsubsection{Mean comparison}\label{s321}

\begin{figure}
\begin{tabular}{@{}cc@{}}
(a)&(b)\\

\includegraphics{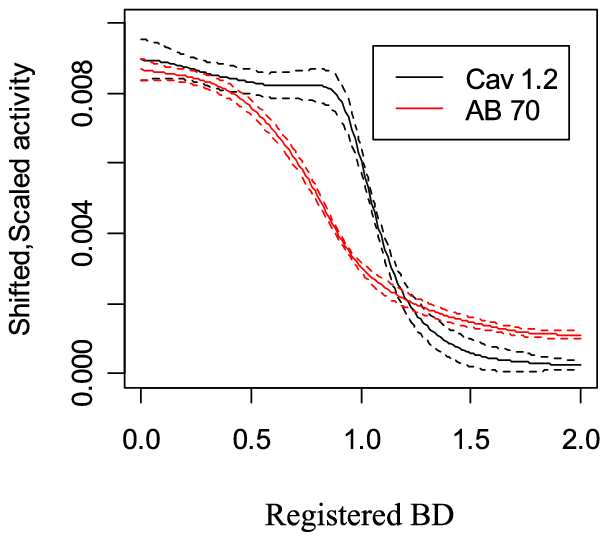}
&\includegraphics{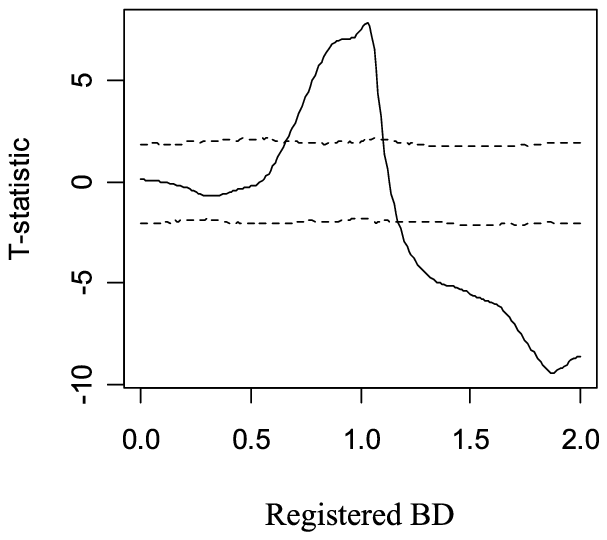}\\
(c)&(d)\\

\includegraphics{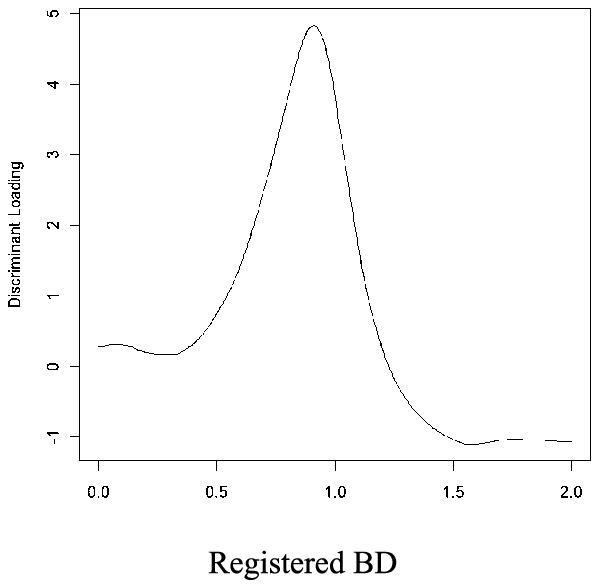}
&
\includegraphics{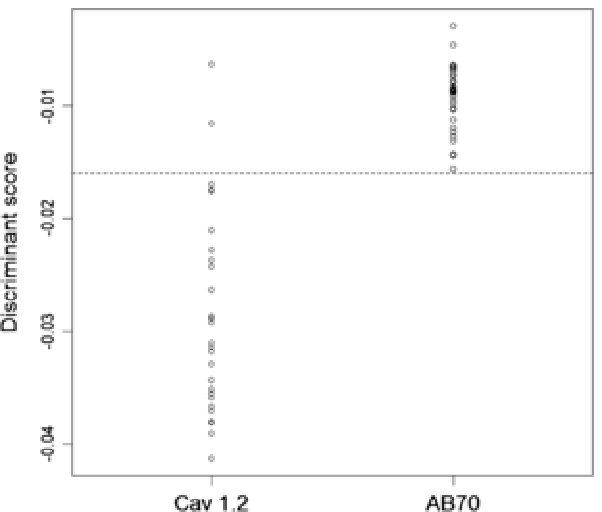}

\end{tabular}
  \caption{\textup{(a)} Comparison of mean (scaled and registered) expression
curves for Cav1.2 (black) and AB70 (red) expression. Curves are
obtained by pointwise averaging across nuclei within each image. Dashed
lines show pointwise 95\% confidence intervals. \textup{(b)} Solid line is
curve of test statistic for two sample T-test, calculated pointwise,
between Cav1.2 and AB70 expression curves shown in \textup{(a)}.
Dashed lines denote 95\% confidence band for test statistic, generated
by repeated randomization. \textup{(c)} Plot of coefficients of the
discriminant function after regularized linear discriminant analysis
between Cav1.2 and AB70 expression curves shown in Figure \protect\ref{f4}\textup{(d)}.
\textup{(d)} Plot of discriminant scores by group with dashed line showing
optimal threshold.}\label{f5}
\end{figure}

Comparison of group means (calculated pointwise) shows slightly
different profiles for Cav1.2 and AB70 [Figure \ref{f5}(a)]. Cav1.2 expression
appears to remain constant up to the cell boundary, whereupon there is a
sharp drop-off, with little expression beyond the boundary. For AB70,
the drop-off is more gradual and some expression appears to extend
beyond the boundary. We tested the null hypothesis $H_0: \mu_{A} =
\mu_{C}$, against a general alternative using $T(r_{i})=(\hat{\mu}_{C}(r_{i})-\hat{\mu}_{A}(\delta_{A}r_{i}))(n_{C}^{-
1}s_{C}^{2}(r_{i})+n_{A}^{-1}s_{A}^{2}(\delta_{A}r_{i}))^{-
0.5}$, $i=1,2,\ldots,200$. Here $s_{C}^{2}(r_{i})$ is the sample
variance of the (registered) expression distributions for the Cav1.2
group at $r_{i}$ and $n_{C}=27$ is the number of cells in this group.
Similar notation is used for the AB70 group, with $n_{A}=38$. The
significance of the test statistic was computed by means of repeated
randomization: under the null hypothesis, two sets of expression
distributions of size $n_{A}$ and $n_{C}$ were repeatedly randomly
sampled (${N}=5000$ times) without replacement from the combined
collection of $n_{A}+n_{C}$ pooled expression distributions from both
groups. For each sample (permutation), ${\operatorname{sup}}|T(r)|$ was computed.
Approximate 95\% simultaneous critical levels were computed as $\pm T_{0.975} = 2.45$, which is the 97.5th percentile of the
${\operatorname{sup}}|T(r)|$ statistics across permutations. Although the test statistic
does not appear to be significant near the center of the nuclei ($r=0)$, we see that the observed test statistic is above the confidence
band in a region close to the boundary ($r=1)$, while it is below the
confidence band outside the nucleus boundary [Figure \ref{f5}(b)].

\subsubsection{Penalized discriminant analysis}\label{s322}

To further describe the difference between the groups, we consider the
problem of discriminating between the groups using the average
expression curves $g_{k}$ using Fisher's linear discriminant analysis
(LDA). The discriminant $\mathbf{d=W}^{\mathbf{-1}}(\hat{\bolds{\mu}}_{\mathbf{A}}-\hat{\bolds{\mu}}_{\mathbf{C}})$,
where $\mathbf{W}$ is the within class covariance
matrix, $\hat{\bolds{\mu}}_{\mathbf{C}}=( \hat{\mu}_{C}( r_{1}),\ldots,\hat{\mu}_{C}( r_{200}))$, $\hat{\bolds{\mu}}_{\mathbf{A}}=(
\hat{\mu} _{A}( \delta_{A}r_{1} ),\ldots,\hat{\mu} _{A}( \delta_{A}r_{200}))$. In classical statistics, \textbf{W} is estimated by the pooled
sample variance covariance matrix, that is, $\mathbf{W}=0.5\bolds{\Sigma}^{\mathbf{A}} + 0.5\bolds{\Sigma}^{\mathbf{C}}$, where $\bolds{\Sigma}
_{ij}^{A}=(27-1)^{-1}\sum_{k}(\hat{g}_{k}^{A}(r_{i})-\hat{\mu}_{A}(r_{i}))(\hat{g}_{k}^{A}( r_{j})-\hat{\mu}_{A}(
r_{j} ) )$, $i,j =1,\ldots,200$, is the sample within group
variance covariance matrix for the AB70 group and $\bolds{\Sigma}^{\mathbf{C}}$
is similarly defined for the Cav1.2 group [\citet{anderson}]. However,
the dimension of the expression curves (200) exceeds sample size (65),
causing $\mathbf{W}$ to become singular, which causes problems when
computing $\mathbf{d}$. To ensure stable inversion, we instead compute a
penalized within class covariance matrix $\mathbf{W}_{p}=
0.5\bolds{\Sigma}^{\mathbf{A}} + 0.5\bolds{\Sigma}^{\mathbf{C}} + \lambda\mathbf{I}$,
where $\mathbf{I}$ is a $200\times 200$ identity matrix and $\lambda$ is a
regularization parameter. We use the decision rule: nucleus $k$ belongs
to AB70 if $\mathbf{d}_{P}^{T}\mathbf{\hat{g}}_{k} > \tau$,
where $\mathbf{d}_{p}=\mathbf{W}_{p}^{ - 1}( \hat{\bolds{\mu}}
_{\mathbf{A}}-\hat{\bolds{\mu}} _{\mathbf{C}} )$is the penalized
discriminant and $\tau$ is a predetermined threshold. A leave out one
cross-validation (CV) procedure is used to choose a combination of
$\lambda$ and $\tau$ which jointly minimize misclassification error
[\citet{hastie}]. Using grid search over a range
of $\lambda$ (10 values between 0.0001 and 0.1 equispaced on a
logarithmic scale) and $\tau$ (10 equispaced values between 0.5 and
1.5), a unique minimum CV error of 2 misclassifications out of 65 (i.e.,
3\%) was obtained for $\lambda = 0.0007$ and $\tau = 1.17$ [Figure \ref{f5}(d)].
The optimal penalized discriminant is practically zero from the center
out, has a sharp dip near the boundary and an elevated level beyond it
[Figure \ref{f5}(c)]. The use of a Laplacian type penalty, suggested by
[\citet{friedman}], instead of $\mathbf{I}$ does not appear to produce a
well conditioned matrix $\mathbf{W}_{p}$ in this case.

\subsection{Flexible parametric modeling}\label{s33}

The analysis of the previous section has demonstrated some differences
in the average expression of the two types of VGCC makers near the
boundary of the nucleus and possibly beyond. In this section we attempt
to better characterize these differences by fitting separate linear
models to the expression in three regions: the interior of the nucleus,
the nuclear boundary and the exterior, using a piecewise linear model
\begin{equation}\label{e34}
\mathit{g}^{P}( r )=\mathrm{E}[ h( a|r ) ] = \sum_{i = 1}^{3} a_{i} +
b_{i}rI\bigl\{\kappa_{i}<r<\kappa_{(i+1)}\bigr\}.
\end{equation}
Here $\kappa = \{\kappa_{i}, i = 1,2,3,4\}$ are knot points with
$\kappa_{1} = 0$ and $\kappa_{4} = 2$, $\mathbf{a} = (a_{1}, a_{2},
a_{3})$ are intercepts and $\mathbf{b} = (b_{1}, b_{2}, b_{3})$ are
slopes. The knotpoints $\kappa_{2}$ and $\kappa_{3}$ allow flexibility
in choosing the extent of the `boundary' region. Unlike usual
implementations of piecewise models, model (\ref{e34}) does not impose continuity
across knots. This allows parameters for each piece to be estimated
mutually independently, simplifying inference. We use the weighted least
squares criterion $L( \kappa,\mathbf{a},\mathbf{b} ) = \sum w( r )( h( a|r ) -
g^{P}( r ) )^{2}$ for model fitting. The weighting function $w(r)$ is the
same as used in (\ref{e33}), to account for sampling density. For given knot
points, $\kappa_{2}, \kappa_{3}$, the criterion
$L(\kappa,\mathbf{a},\mathbf{b})$
can be fit as three separate linear models using weighted least squares.
However, when two successive pieces have identical slopes and
intercepts, that is, $a_{i} = a_{(i+1)}$ and $b_{i} = b_{(i+1)}$, the
choice of $\kappa_{i}$ is not unique, since any value of $\kappa_{i}$ in
[$\kappa_{(i-1)}, \kappa_{(i+1)}$] will yield the same value of
$L(\kappa,\mathbf{a},\mathbf{b})$. To avoid this ambiguity, we instead minimize a
penalized weighted least squares criterion of the form
\begin{equation}\label{e35}
\qquad L_{p}( \kappa,\mathbf{a},\mathbf{b})=\sum_{r}w(r)\bigl(h(a|r)-g^{P}(r)\bigr)^{2}+\lambda P(\kappa_{2}-1)+\lambda P(1-\kappa_{3}).
\end{equation}
Here $P$ is an asymmetric penalty function: $P(x)=\infty$ if $x\geq 0$ and $P(x)=x^{2}$ if $x<0$. We note that $P$ is a penalty on knot
location, quite different from the smoothness penalty commonly used in
function estimation [\citet{hastie}]. It enables the pieces to be
kept on the correct side of the nuclear boundary and the boundary piece
to be relatively short. To obtain the minimizer of (\ref{e35}), we adopt a two-stage procedure:

\textit{Step 1}: For given $\kappa_{2}, \kappa_{3}$, we compute the
minimizer of $L(\kappa,\mathbf{a},\mathbf{b})$ by weighted least squares
as $\mathbf{\hat{a}}_{\kappa},\mathbf{\hat{b}}_{\kappa}$. These are
computed across a triangular grid of knot points $\kappa_{2jm} =
0.01j$, $\kappa_{3jm} =0.01m$, $j = 1,\ldots, 100$, $m = j,\ldots,
100$.

\textit{Step 2}: The criterion $L_{p}( \kappa,\mathbf{\hat{a}}_{\kappa},
\mathbf{\hat{b}}_{\kappa} )$ is computed for all knot points in the
grid of $\kappa$ values using (\ref{e35}). The regularization parameter $\lambda$
is chosen by grid search to be the smallest value which ensures unique
estimation of knot points. The global minimum of $L_{p}(
\kappa,\mathbf{a}_{\kappa}, \mathbf{b}_{\kappa} )$ is obtained by grid
search over $\kappa$ values.

\begin{figure}
\begin{tabular}{@{}cc@{}}
(a)&(b)\\

\includegraphics{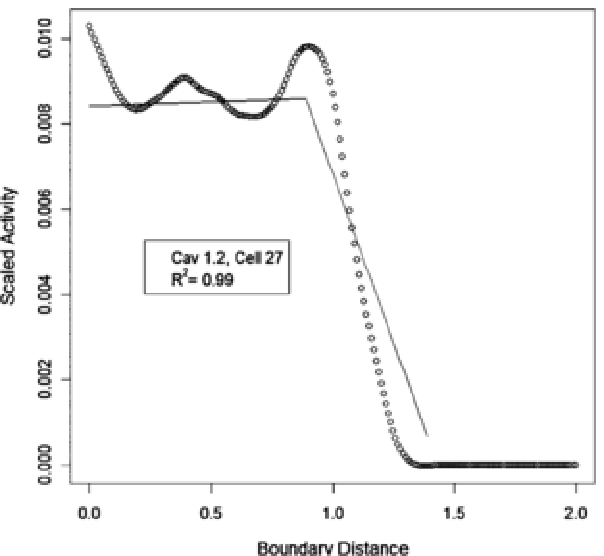}
&
\includegraphics{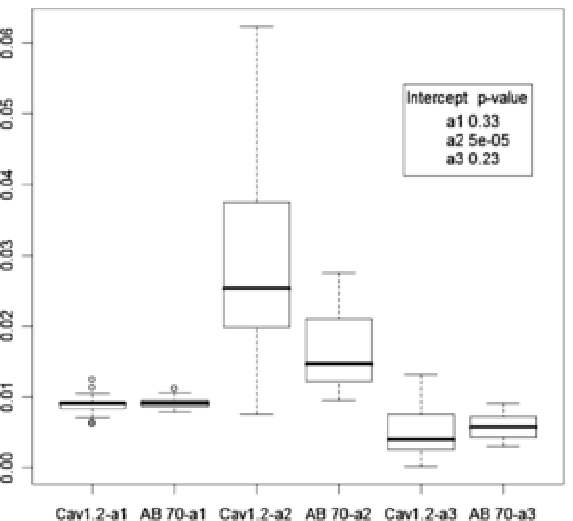}
\\
(c)&(d)\\

\includegraphics{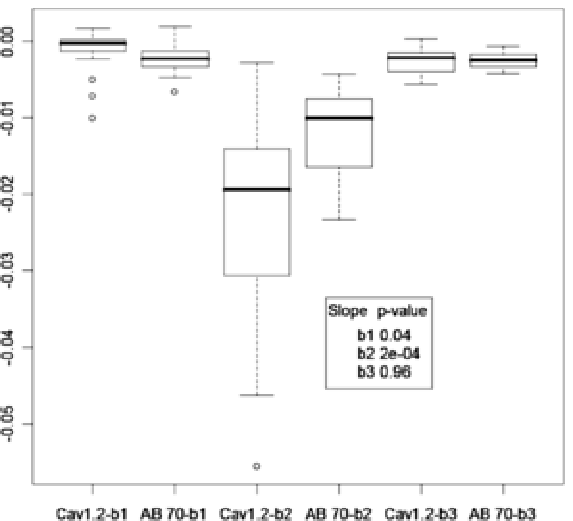}
&
\includegraphics{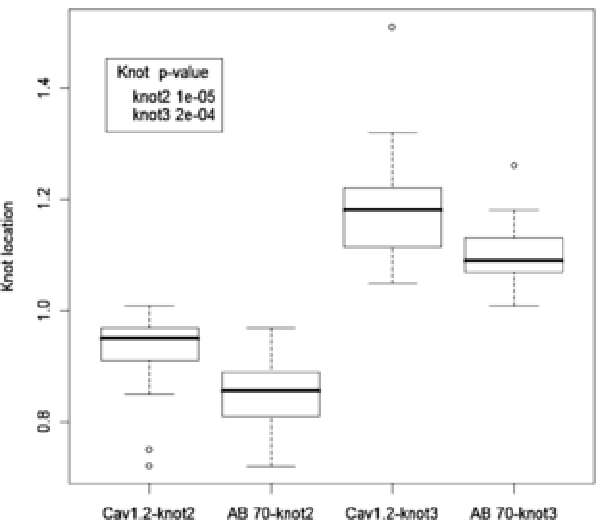}

\end{tabular}
  \caption{\textup{(a)} Example of adaptive piecewise linear fit to a scaled
Cav1.2 expression curve. Dots show actual curve and lines show
three piece linear fit. \textup{(b)}--\textup{(d)} Subject specific piecewise linear
modeling with p-values from two sample \textit{t}-tests comparing parameter
estimates for Cav1.2 and AB70 curves: \textup{(b)} intercepts,
\textup{(c)}
slopes, \textup{(d)} knot locations.}\label{f6}
  \end{figure}

Fits from piecewise modeling closely match the average expression curves
obtained by spline smoothing [Figure \ref{f7}(a)], with median $R^{2}( = 1 -
\sum( \hat{g}_{k} - \hat{g}_{k}^{P} ) / V( \hat{g}_{k} ) )$ values of
0.99 for both AB70 and Cav1.2 groups.

Comparison of intercepts across groups shows no significant differences
in any of the three regions [Figure \ref{f6}(b)]. Comparison of slopes of each
piece across groups shows no significant difference for the last piece,
which represents expression beyond the nuclear boundary [Figure \ref{f6}(c)].
The difference in the first pieces near the center of the nucleus is
marginally statistically significant, but with very little absolute
change in median slope value. The main difference between the Cav1.2 and
AB70 groups lies in the middle piece (across the boundary), with Cav1.2
having significantly lower (steeper) slopes. Cav1.2 knots also appear to
occur later than AB70 knots on average [Figure \ref{f6}(d)].

\section{Paired analysis of radial maps}\label{s4}

\begin{figure}
\begin{tabular}{@{}cc@{}}
(a)&(b)\\

\includegraphics{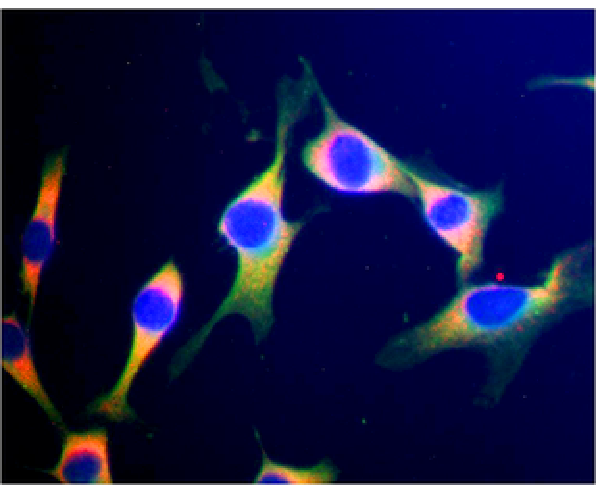}
&
\includegraphics{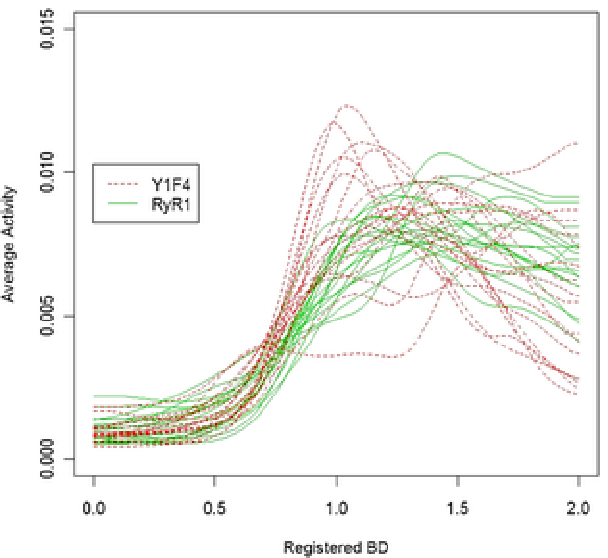}
\\
(c)&(d)\\

\includegraphics{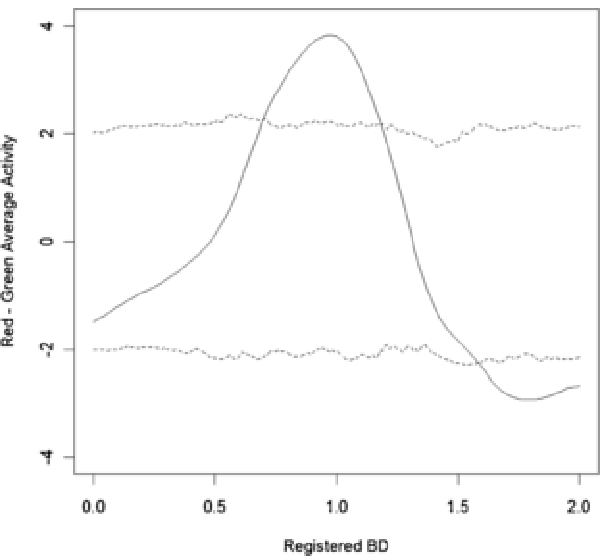}
&
\includegraphics{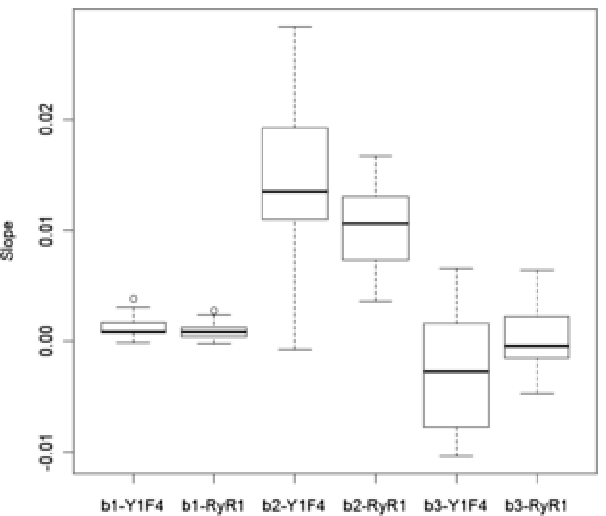}

\end{tabular}
   \caption{\textup{(a)} Optical fluorescence microscopy image of JEG-3 trophoblastic
cells. Images are labeled in a blue chromatin marker (`DAPI'), a red
marker for all forms of sarcoplasmic/endoplasmic reticulum calcium
ATPase (`Y1F4') and a green marker [the type 1 ryanodine receptor
(`RyR1')]. \textup{(b)} Comparison of scaled and registered average expression
curves of Y1F4 and RyR1. \textup{(c)} Solid line is curve of test statistic for
paired \textit{T}-test, calculated pointwise, between Y1F4 and RyR1 for
expression curves shown in \textup{(b)}. Dashed lines denote 95\% confidence band
for test statistic, generated by restricted randomization. \textup{(d)}
Comparison of estimated slopes for Y1F4 and RyR1 from piecewise linear
modeling.}\label{f7}
  \end{figure}

In the first example [Figures \ref{f1}(a) and (b)], we compared radial
distributions of markers (Cav1.2 and AB70) from different cells. In the
second example [Figure \ref{f7}(a)], we are interested in comparing two markers
(Y1F4 and RyR1) present in the same cell. From a statistical
perspective, the first example is a two sample problem, whereas the
second example is a paired sample one. In the second example, construct
BD maps for individual cells as described in Section \ref{s2}. Subsequently, we
compute average expression curves for each marker as described in
Section \ref{s31} and then scale them as described in Section \ref{s311}. A simpler
procedure is required for curve alignment, since the pairs of curves
within each cell are automatically aligned. We construct a paired
registration criterion $\mathit{WREGSSEP}$, where $Y$ and $R$ denote Y1F4
and RyR1 respectively:
\begin{eqnarray}\label{e41}
\mathit{WREGSSEP}
&=&
\sum_{k=1}^{nc}\int_{0}^{2}w(r)\bigl\{\bigl(g_{k}^{Y}(r\delta_{k})-\mu^{Y}(r)\bigr)^{2}\nonumber
\\[-8pt]\\[-8pt]
&&\hspace*{55pt}{}+
\bigl( g_{k}^{R}( r\delta_{k} )-\mu^{R}(r)\bigr)^{2}\bigr\}\,dr.\nonumber
\end{eqnarray}
The other terms are as in (\ref{e33}). Minimization of $\mathit{WREGSSP}$ and
subsequent curve alignment were accomplished for $nc = 17$ curves
(from two images) using the iterative algorithm described in Section
\ref{s312}.

From Figure \ref{f7}(b), we can see that the RyR1 (green) curves display a
coherent pattern: their intensity peaks somewhere beyond the nuclear
boundary. Thereafter, their expression remains constant. For the Y1F4
(red) curves, there appear to be two subpopulations. One subpopulation
peaks at the nuclear boundary, followed by a sharp decline in average
intensity. The other subpopulation plateaus near the nuclear boundary,
but then their average intensity increases with increasing radial
distance. A paired \textit{t}-test between the two populations [Figure \ref{f7}(c)]
shows significant difference in average expression between the two
makers near the nuclear boundary and at points beyond a distance of 1.6.
Confidence bands for the paired test statistic were computed using
restricted randomization, that is, each pair of Y1F4 and RyR1 average
expression curves was randomly reassigned to one of two groups each. The
null distribution of the test statistic was then approximated using the
procedure described in Section \ref{s321}. The first penalized linear
discriminant (not shown) is similar in shape to the paired \textit{T}-statistic.
A minimum CV misclassification error rate of 47\% was obtained for this
data set (Section \ref{s322}). Finally, the piecewise linear model (\ref{e34}) was fitted
to individual expression curves. The quality of fit was typically very
good (median ${R}^{2}$ of 0.98). Primary interest lies in the intensity
gradient (slope) for the third part, which is beyond the nuclear
boundary. For RyR1, we see a tight slopes distribution centerd at 0
[Figure \ref{f7}(d)]. For Y1F4, we see a more dispersed slope distribution,
with a preponderance of negative slopes. A paired \textit{t}-test of mean slope
difference shows a significant difference (\textit{p}-value 0.004) between the
markers.

\section{Discussion}\label{s5}

We have presented a modern statistical approach for the analysis of
marker expression distributions under boundary distance mapping. The
technical improvements proposed include the following: (i) Extension of
the Euclidean distance map to points outside the boundary. (ii)
Presmoothing and oversampling of object boundaries for improved
estimation of boundary distances. (iii) Variable bandwidth smoothing of
marker expression distributions. (iv) Scaling and shifting of average
expression curves to account for variations in lighting and incorrect
boundary identification. (v) Comparison of average expression curves
across experimental conditions using suprema of \textit{t}-tests and penalized
discriminant analysis. (vi) Targeted inference on regionwise group
differences by flexible parametric modeling. The methods are illustrated
using two experiments involving calcium channels, however, the proposed
techniques are general enough to be immediately applicable to other
types of experiments, for example, the study of chromatin structure
[\citet{bewersdorf}] or other nuclear proteins [\citet{knowles}]. In order to be applicable at larger scales, however, automated
methods of image segmentation are required, for example, to identify
nuclear/cellular boundaries. The success of automated techniques
typically varies, depending on the quality/resolution of imaging as well
as the complexity of the field of view [\citet{jahne}]. In the future we
also hope to extend boundary distance analysis to more complex features,
such as the local structure of marker expression.

The main findings of the analysis of the VGCC experiment are that Cav1.2
appears to have a uniform distribution throughout the nucleus which
vanishes outside the nuclear boundary. Conversely, the expression of
AB70 appears to gradually decrease as it reaches the periphery of the
nucleus and some expression appears to persist beyond the nuclear
boundary. The relatively clean separation between these two proteins
(misclassification error rate of 3\%) may indicate that there is a
difference in transmembrane function of the channel proteins recognized
by the antibodies Cav1.2 and AB70. The functional consequences of these
differences will be the subject of future investigations.

In the JEG-3 cell-line experiment, differences in the distribution of
RyR1 and SERCA (Y1F4) are not that clear (misclassification error rate
of 47\%). This is not unexpected, since both proteins would be expected
to be located in the ER of these trophoblasts. Our finding of
heterogeneity in the Y1F4 average expression curves suggests that only
certain subdomains of the ER within JEG-3 cells could be specialized for
SERCA-mediated Ca$^{2+}$ uptake. This possibility is not without
precedent, since the SR of striated muscle is functionally and
anatomically divided into subdomains specialized for either Ca$^{2+}$
uptake or for Ca$^{2+}$ release [\citet{mackrill}]. In early video
microscopy studies using Ca$^{2+}$-sensitive fluorophores it was noted
that sister cells displayed distinct Ca$^{2+}$ responses to hormonal
stimulation [\citet{ambler}]. Epigenetic variations in the
abundance (intensity) of Ca$^{2+}$-signalling components between
individual cells in a population could give rise to such differences.

We have proposed a modification of the Euclidean boundary distances
[\citet{knowles}] to measure boundary distance for points outside
the object boundary. A similar extension for erosion-based distance
measurement [\citet{bewersdorf}], using dilation instead of
erosion, is straightforward [\citet{bewersdorf}]. Similarly, the
methodology described here can extend to 3-d stacks of images in a
straightforward manner. However, we note that the methodology described
here can be satisfactorily applied only in situations where the
orientation of expression/objects is not of interest, since all
orientation information is lost in the profile distributions.

The attraction of the FDA approach lies in the fact that it extends
standard univariate statistical techniques like ANOVA and \textit{t}-tests to
curve data [\citet{ramsay}]. However, the necessity of
preprocessing curves by registration can mean that some information
about differences between groups can be lost. The adaptive piecewise
linear approach proposed in Section \ref{s33} avoids this loss of information.
Significant differences in the distribution of knot points across groups
indicate that this may indeed be the case. Moreover, piecewise linear
modeling also reveals that the difference in the average expression
curves may not be in their magnitude, but in their slopes.

\section*{Acknowledgments} The first author
would like to thank Trevor Hastie and Rob Tibshirani for helpful
discussions.

\printaddresses

\end{document}